\documentclass[prb,groupedaddress,twocolumn,showpacs]{revtex4-1}

\usepackage{graphicx}
\usepackage{dcolumn}
\usepackage{bm}
\usepackage{hyperref}
\usepackage{url}
\usepackage{color}

\newcommand{\pft}{PbFe$_{1/2}$Ta$_{1/2}$O$_{3}$\,}
\newcommand{\pfts}{PFT\,}
\newcommand{\pfn}{PbFe$_{1/2}$Nb$_{1/2}$O$_{3}$\,}
\newcommand{\pfns}{PFN\,}
\definecolor{reddeep}{rgb}{0.995,0.000,0.200}

\begin{document}
\title{Magnetic short and long range order in \pft}

\author{S. Chillal}
\email[]{chillals@phys.ethz.ch}
\affiliation{Neutron Scattering and Magnetism, Laboratory for Solid State Physics, ETH Z\"urich, Z\"urich, Switzerland}

\author{S. N. Gvasaliya}
\affiliation{Neutron Scattering and Magnetism, Laboratory for Solid State Physics, ETH Z\"urich, Z\"urich, Switzerland}

\author{D. Schroeter}
\affiliation{Institut f\"{u}r Physik der Kondensierten Materie, Technische Universit\"at Braunschweig, 38106 Braunschweig, Germany}

\author{M. Kraken}
\affiliation{Institut f\"{u}r Physik der Kondensierten Materie, Technische Universit\"at Braunschweig, 38106 Braunschweig, Germany}

\author{F. J. Litterst}
\affiliation{Institut f\"{u}r Physik der Kondensierten Materie, Technische Universit\"at Braunschweig, 38106 Braunschweig, Germany}

\author{T. A. Shaplygina}
\affiliation{Ioffe Physico-Technical Institute RAS, 194021, St. Petersburg, Russia}

\author{S. G. Lushnikov}
\affiliation{Ioffe Physico-Technical Institute RAS, 194021, St. Petersburg, Russia}

\author{A. Zheludev}
\affiliation{Neutron Scattering and Magnetism, Laboratory for Solid State Physics, ETH Z\"urich, Z\"urich, Switzerland}

\begin{abstract}
\pft belongs to the family of PbB$'_{x}$B$''_{1-x}$O$_{3}$ which have inherent chemical disorder at the B-site. Due to this disorder, a complex magnetic phase diagram is expected in the material. In this paper, we report experimental results of magnetic properties of \pfts through macroscopic characterization, neutron scattering and M\"{o}ssbauer spectroscopy techniques. With these results we show for the first time that \pft behaves very similar to \pfn, i.e, it undergoes AF transition at 153~K and has a spinglass transition at 10~K, below which the antiferromagnetism coexists with spinglass. We suggest that the mechanism which is responsible for such a non-trivial ground state can be explained by a speromagnet-like spin arrangement similar to the one proposed for \pfn.   
\end{abstract}
\pacs{{75.50.Ee} {Antiferromagnetic materials}; {75.50.Lk} {Spin glasses}; {76.80.+y} {M\"ossbauer spectroscopy of solids}; {28.20.Cz} {Neutron scattering}
}

\date{\today}

\maketitle

\section{Introduction}

Studies of disordered materials have gained interest due to novel phases and phase transitions they exhibit~\cite{Dagotto2005,Gabay1981,Palmer1984,Viet2009}. AB$'_{x}$B$''_{1-x}$O$_{3}$ are complex perovskites which have inherent disorder at the B-site resulting in unusual properties. In particular, magnetic ion dilution at the disordered B-site leads to rich magnetic phase diagram and magnetic ground states ranging from simple antiferromagnet to incommensurate structures~\cite{Smolenskii1982}. Furthermore, spinglass properties have also been reported in some of the complex disordered perovskites~\cite{Falqui2005,Choudhury2012,Battle1989}. More recently, our studies of the disordered antiferromagnet \pfn (PFN ) demonstrated that long range antiferromagnetic order can co-exist with a spinglass state on the microscopic scale~\cite{Chillal2013}. Below we refer to this ground state as AFSG.

Coexistence of long-range and spinglass orders have been studied in disordered ferromagnetic materials both theoretically and experimentally~\cite{Gabay1981,Lauer1982,Ryan2003,Ren1995,Ryan2000,Mirebeau1997,Abd1986,Campbell1983}. For example, for {\it reentrant spinglasses} in amorphous metallic compounds Fe-Mn~\cite{Lauer1982,Ryan2003}, Fe-Zr~\cite{Ren1995,Ryan2000}, Au-Fe~\cite{Campbell1983,Abd1986} a spin canting along with transverse spin freezing was considered as a mechanism of coexistence. Random transverse spin freezing scenario for occurrence of such a ground state is predicted by mean-field theory~\cite{Gabay1981}.

On a similar footing we explained a novel AFSG phase in the crystalline disordered perovskite \pfns through canting of magnetic moments of Fe$^{3+}$ culminated in their freezing~\cite{Chillal2013}. What remained unclear is whether the essential ingredients for such spin arrangement is disorder and magnetic dilution themselves or the particular magnetic ion at the disordered site. An extensive study of influence of the non-magnetic ions on the phase diagram of \pfns allows us to better understand magnetic ground states in AB$'_{x}$B$''_{1-x}$O$_{3}$ type complex perovskites. To date, the role played by non-magnetic ions in formation of AFSG state in these perovskites remains unclear. 
Previous studies of \pfns diluted by non-isovalent Ti$^{4+}$ at the Fe/Nb site or by isovalent Ba$^{2+}$ at the Pb site suggested that its magnetic properties are indeed strongly affected by the type of non-magnetic ion in the lattice~\cite{Laguta2013}.
In the present work we make the next logical step and focus on the fully substituted stoichiometric lead iron tantalate, \pft (\pfts), a close relative of \pfns. In this case the Nb$^{5+}$ ion is isovalently substituted by Ta$^{5+}$. The chemical structures of both materials are essentially identical~\cite{Lampis1999,Bonny1997}. Despite the strong structural similarities, magnetic properties of \pfts are not very well understood in contrast to the generally accepted antiferromagnetic ($\sim$143~K) and spinglass (12~K) transitions in \pfns~\cite{Kumar2008,Gelu2009,Kleemann2010,Chillal2013}. One of the challenges with \pfts is that indications of the magnetic transitions in {\it  dc} magnetic susceptibility are strongly sample-dependent. For example, various sources reported AF transitions through anomalies in the {\it dc} magnetization in the range of 130-180~K~\cite{Lampis2000,Shvorneva1966,Nomura1968,Martinez2010}. This transition was confirmed by the appearance of AF Bragg peak in neutron diffraction~\cite{Ivanov2001} suggesting a simple G-type structure with $\sim$3$\mu_{B}$ magnetic moment per Fe$^{3+}$ ion at base temperature~\cite{Ivanov2001}. However, first principle calculations of the electronic structure of \pfts suggested possibility of a second AF transition~\cite{Lampis2004} at 48~K which was claimed to be observed experimentally at $\sim$55~K by {\it Martinez} et.al~\cite{Martinez2010}. In addition, the {\it  dc} susceptibility results by {\it Falqui} et. al showed a maximum in ZFC data around 9~K which exhibits properties of a spinglass transition~\cite{Falqui2005}. Overall, there is uncertainty in the number and types of magnetic phases in \pfts. 

In the present work we seek to sort out the magnetic phase diagram of \pft. We employ bulk magnetization, neutron scattering and M\"ossbauer spectroscopy to elucidate magnetic phase diagram. In order to eliminate metallurgical problems in identifying the temperatures of magnetic  transitions, we studied ceramic and single crystal specimens. Both types of samples show similar transition temperatures. Our results prove that  \pfts undergoes two magnetic phase transitions: an AF transition below T$_{N}\sim$153~K and, a SG transition below T$_{SG}\sim$10~K similar to the one observed by {\it Falqui} et al~\cite{Falqui2005}.
Neutron scattering results suggest appearance of short-range magnetic correlations contributing to SG already below $\sim$50~K. Combining the AF Bragg peak observed by neutron scattering with M\"ossbauer spectroscopy we show that the magnetic state of \pfts below T$_{SG}$ is a {\it microscopically coexisting antiferromagnetic spinglass phase}. These results, therefore, enable us to state that the magnetic phase diagram of \pfts is identical to \pfns despite having a different non-magnetic ion sharing Fe$^{3+}$.

\section{Samples \& Experimental Methods}

\begin{figure}
\includegraphics[width=0.99\columnwidth]{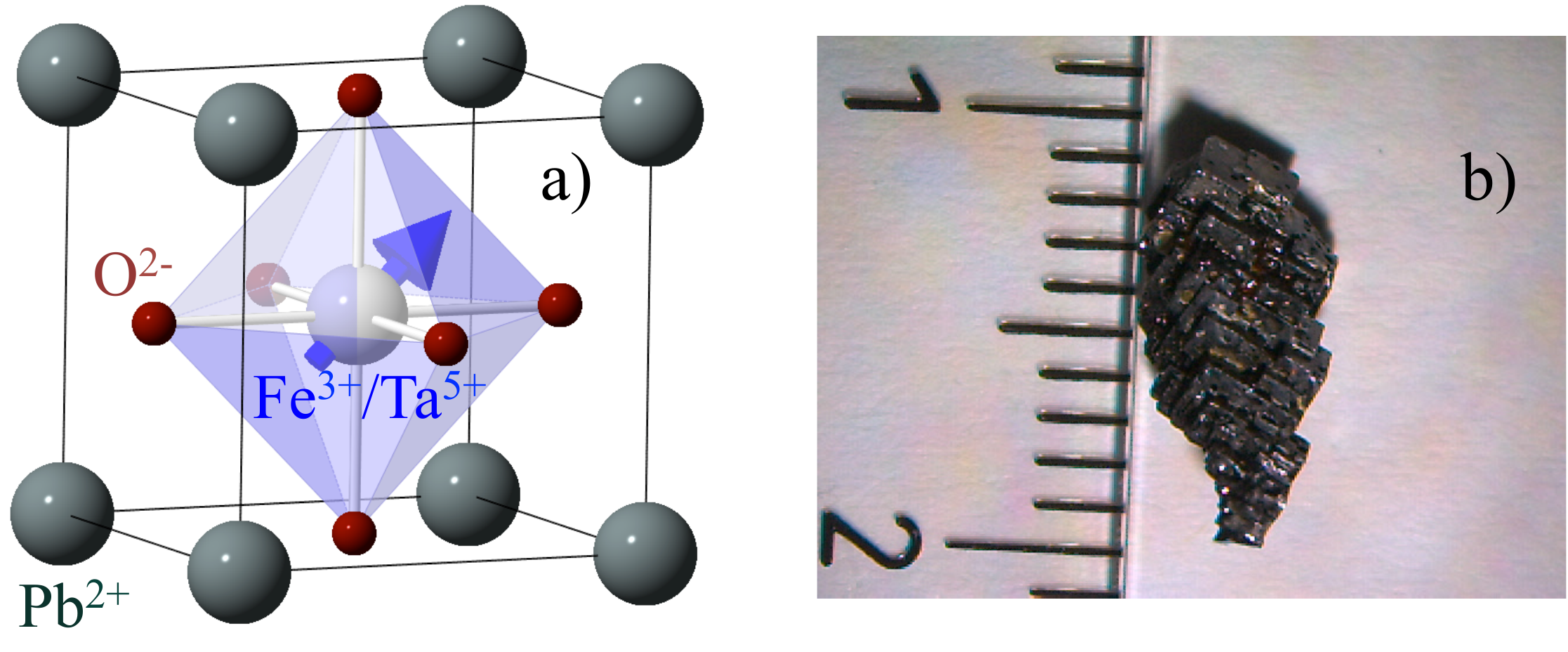}
\caption{a) Perovskite unit cell of \pft ; b) a photo of \pfts single crystal (1x0.5x0.3 cm$^{3}$) used in the neutron scattering experiments.
        } 
\label{Fig:Fig1}
  \end{figure}

\pfts crystallizes in a cubic perovskite structure with Pb$^{2+}$ ions residing at corners of the unit cell, while oxygen octahedra surrounds 
Fe$^{3+}$/Ta$^{5+}$ ions as shown in Fig.~\ref{Fig:Fig1}a. The Fe$^{3+}$  and Ta$^{5+}$ ions are believed to be randomly distributed over 
the B-site of the lattice. Upon lowering the temperature \pfts undergoes several structural phase transitions. However, the distortions involved are rather insignificant~\cite{Lampis2000, Smolenskii1984}.

The \pfts samples used in the present study were synthesized with the same procedure as employed for \pfns~\cite{Chillal2013, Kumar2008}. The  
essential difference is that the sintering temperatures for \pfts are 1100~$^{\circ}$C, 
1250~$^{\circ}$C for ceramics and single crystals respectively.

Macroscopic measurements were made on ceramics and single crystals of \pfts using a Physical Property Measurement System (PPMS). 
Conventional M\"{o}ssbauer absorption spectroscopy experiments were conducted with the help of He-flow cryostat in transmission geometry on 
a powder of \pfts with natural abundance of $^{57}$Fe. 
Neutron scattering experiments were performed at the cold 3-axis spectrometer TASP (SINQ, Switzerland). 
A high-quality single crystal of \pfts (see Fig.~\ref{Fig:Fig1}b) was  aligned in the $<hh0>/<00l>$ scattering plane in cubic notation so that the AF Bragg peak at 
the $\mathbf Q_{AF} = (\frac{1}{2}, \frac{1}{2}, \frac{1}{2})$~\cite{Lampis2000} position is reachable. Most of the neutron data were collected using $k_{f}=1.55~\AA^{-1}$ 
and a collimation of {\it open-80$'$-sample-80$'$-80$'$}. Higher resolution results were obtained with collimation of {\it open-20$'$-sample-20$'$-20$'$}. A liquid nitrogen cooled Be-filter 
was used to suppress higher order contaminations. For polarized neutron diffraction, the MuPAD setup~\cite{Janoschek2007} was employed with a spectrometer configuration $k_{f}=1.97~\AA^{-1}$, leading to effective 
collimation {\it open-80$'$-sample-80$'$-open}. All the neutron data are analyzed by convoluting the respective scattering function with resolution of the spectrometer using {\it ResLib} package~\cite{Zhelud2007}. The calculated resolution reproduces the measured Bragg peaks (110) and $(\frac{1}{2}, \frac{1}{2}, \frac{1}{2})$ leading to effective sample mosaicity values of 12$'$ and 20$'$ respectively in 20$'$-collimation and 80$'$-collimation setups. The effective mosaicity that is obtained for polarized data is 27$'$. 

\section{Results}

\subsection{Bulk Magnetization}

\begin{figure}
\includegraphics[width=0.99 \columnwidth]{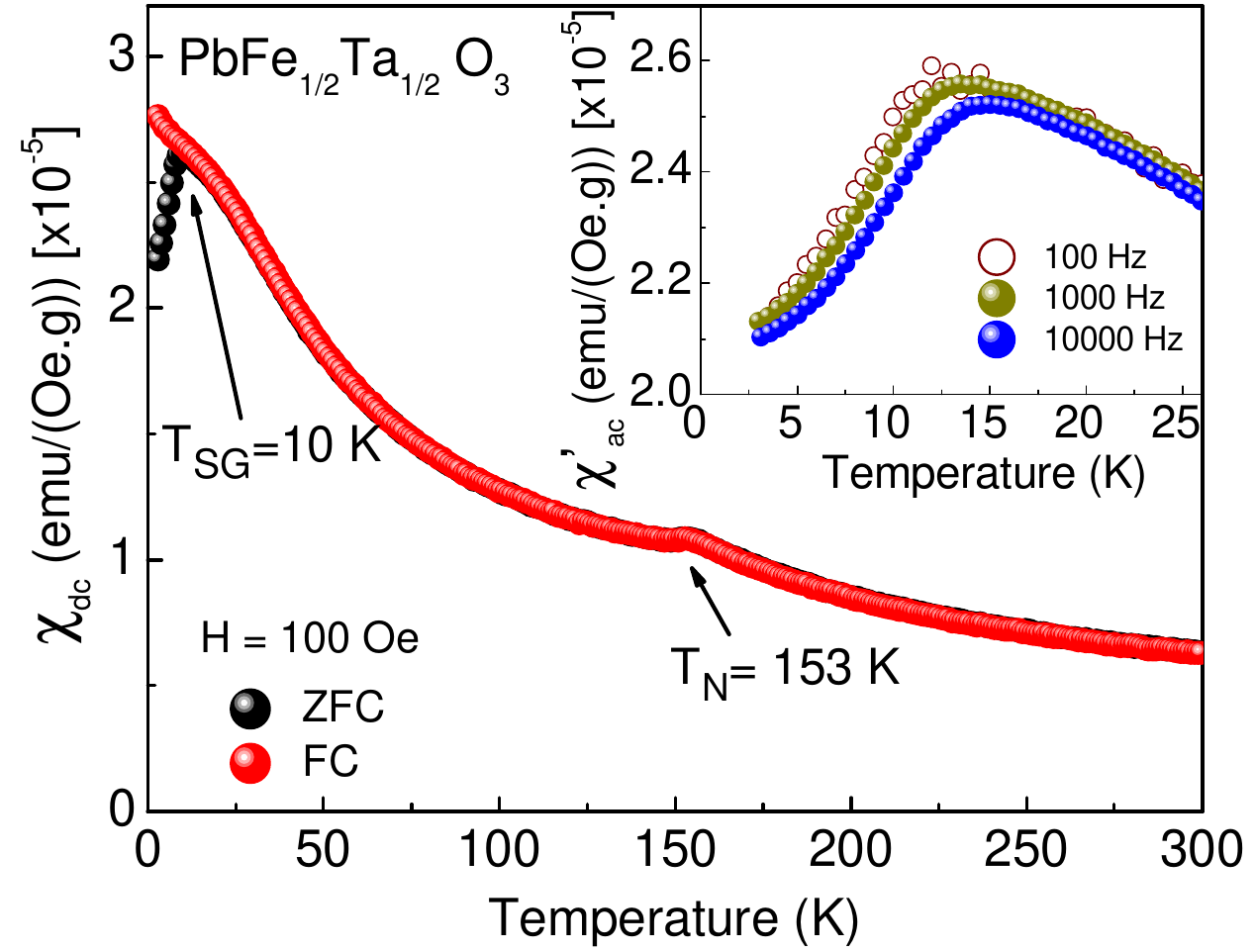}
\caption{ {\it dc} susceptibility of \pfts measured in a field of 100~Oe in standard ZFC and FC protocols for a ceramic sample. Inset shows frequency dependence of the broad peak referring to SG transition in {\it ac} susceptibility. }
  \label{Fig:Fig2}
  \end{figure}

\noindent Two magnetic phase transitions of \pfts are immediately detected in macroscopic experiments. Fig.~\ref{Fig:Fig2} shows the {\it dc} magnetic susceptibility of \pfts powder taken in field-cooled (FC) and 
zero-field-cooled (ZFC) protocols as a function of temperature. At T$_{N}\sim$153~K, \pfts undergoes paramagnetic (PM) to AF transition~\cite{Lampis2000,Shvorneva1966,Nomura1968,Martinez2010,Ivanov2001}, while the 
low-temperature anomaly seen as a splitting of ZFC and FC data around 10 K suggests a second transition from AF into SG phase~\cite{Falqui2005, Martinez2010}. The SG nature of this transition is further verified by 
gradual frequency dependence of {\it ac} susceptibility. Inset of Fig.~\ref{Fig:Fig2} shows a rounded peak around T$_{SG}$ whose maximum shifts to higher temperatures upon increasing the field 
frequency. Additional measurements performed on collections of small single crystals of \pfts give essentially the same results with T$_{N}\sim$158~K 
and T$_{SG}\sim$9~K.  So we conclude that the \pfts samples with very different metallurgy have similar macroscopic properties. 

Macroscopic methods alone are insufficient to properly explore the magnetic phases of \pfts. The development of short range magnetic correlations and the very presence of AF long-range order (LRO) 
in the SG state is best probed with neutron scattering. 

\subsection{Neutron Scattering}

\begin{figure}
\includegraphics[width=1 \columnwidth]{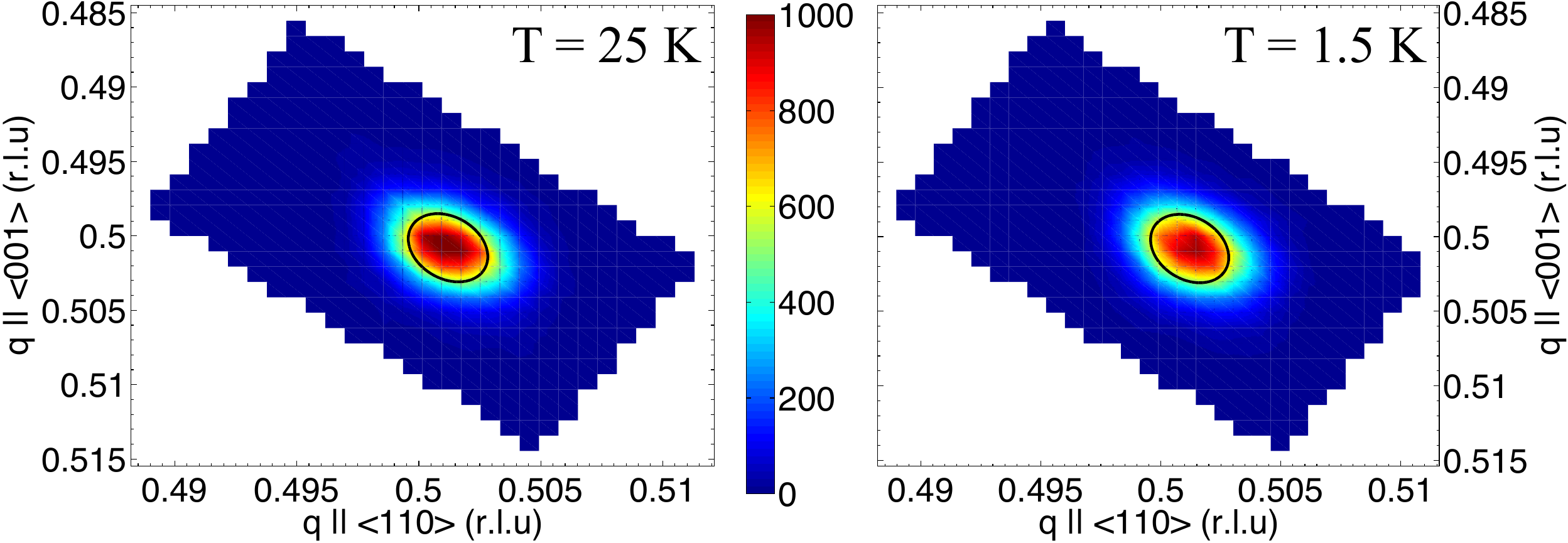}
\caption{False color map of elastic neutron intensity around AF Bragg peak at 25~K and 1.5~K respectively. 
         Weak intensity surrounding the Bragg peak is a contribution from DS, which on this scale of wave vectors appears as (nearly) flat background. 
         The resolution ellipse (black line), calculated from spectrometer parameters matches with the observed contour at half the maximum intensity.}
  \label{Fig:maps}
\end{figure}

As mentioned earlier, \pfts has G-type~\cite{Ivanov2001} long-range AF order which produces a Bragg peak at $\mathbf Q_{AF} = (\frac{1}{2}, \frac{1}{2}, \frac{1}{2})$ position in neutron diffraction. Fig.~\ref{Fig:maps} shows a color map of this Bragg peak measured in a high-resolution set up above and below T$_{SG}$. As depicted in the figure, we observe a sharp resolution limited AF Bragg peak intensity that remains undisturbed while passing through SG transition. Meaning, the true LRO persists into the SG phase.  

\begin{figure}
\includegraphics[width=0.95 \columnwidth]{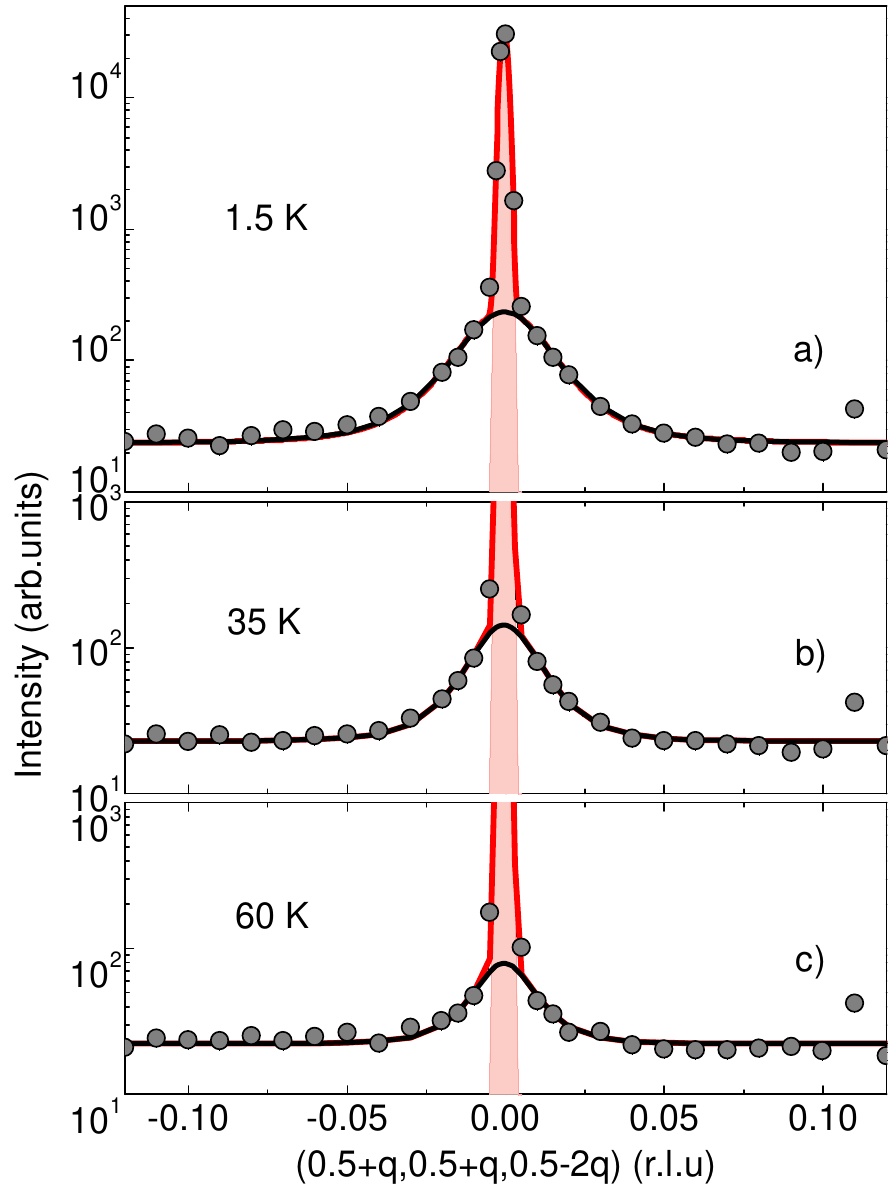}
\caption{Elastic transverse neutron scans around $\mathbf Q_{AF} =(\frac{1}{2}, \frac{1}{2}, \frac{1}{2})$ demonstrating the temperature evolution of AF Bragg and of the diffuse scattering. 
         The data are collected with unpolarized neutrons. Logarithmic scale is used on Y-axis to highlight the relatively weak magnetic DS intensity. Solid lines are fits to the 
         function Eq.~\ref{eq:1}-\ref{eq:3} as described in the text. Red lines correspond to the best-fit results, shaded areas highlight the AF Bragg peak, 
         and the black lines refer to the contribution from DS due to short-range correlations.}
  \label{Fig:spectra1}
\end{figure}

In addition to the LRO, short-range correlations can also be observed by neutron diffraction in the form of diffuse scattering (DS). In our material, these short-range correlations give rise to a broad peak under the sharp AF Bragg reflection as shown in Fig.~\ref{Fig:spectra1}. Relatively big single crystal of \pfts had to be used for these measurements in order to investigate the weak neutron DS signal. As a result the intensities of nuclear and possibly magnetic Bragg peaks suffer from extinction effects. Hence, the temperature evolution of ordered magnetic moment could not be studied in our experiments.

\begin{figure}
\includegraphics[width=0.95 \columnwidth]{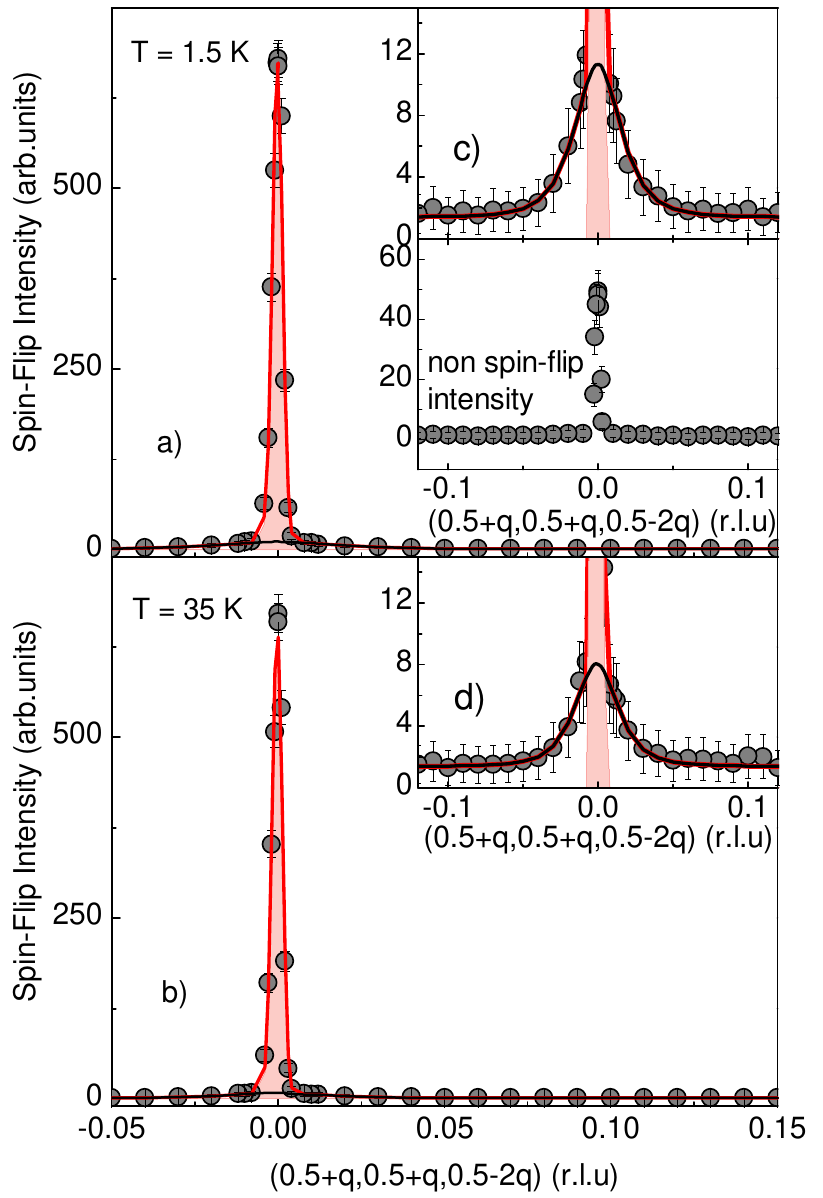}
\caption{Polarized neutron elastic scans from \pfts around $\mathbf Q_{AF} =(\frac{1}{2}, \frac{1}{2}, \frac{1}{2})$ demonstrating magnetic origin of the                            Bragg peaks and DS intensity. i.e, the intensity in non spin-flip channel (inset of a)) is negligible compared to that observed in spin-flip channel (a)). c)-d) highlight the respective DS intensity in a)-b).
         Solid lines are fits to the functions Eq.~\ref{eq:1}-\ref{eq:3} as described in the text. Red lines correspond to the best-fit results, shaded areas refer to AF Bragg peak, 
         and the black lines emphasize contribution from magnetic DS.}
  \label{Fig:spectra2}
\end{figure}

\begin{figure}
\includegraphics[width=1 \columnwidth]{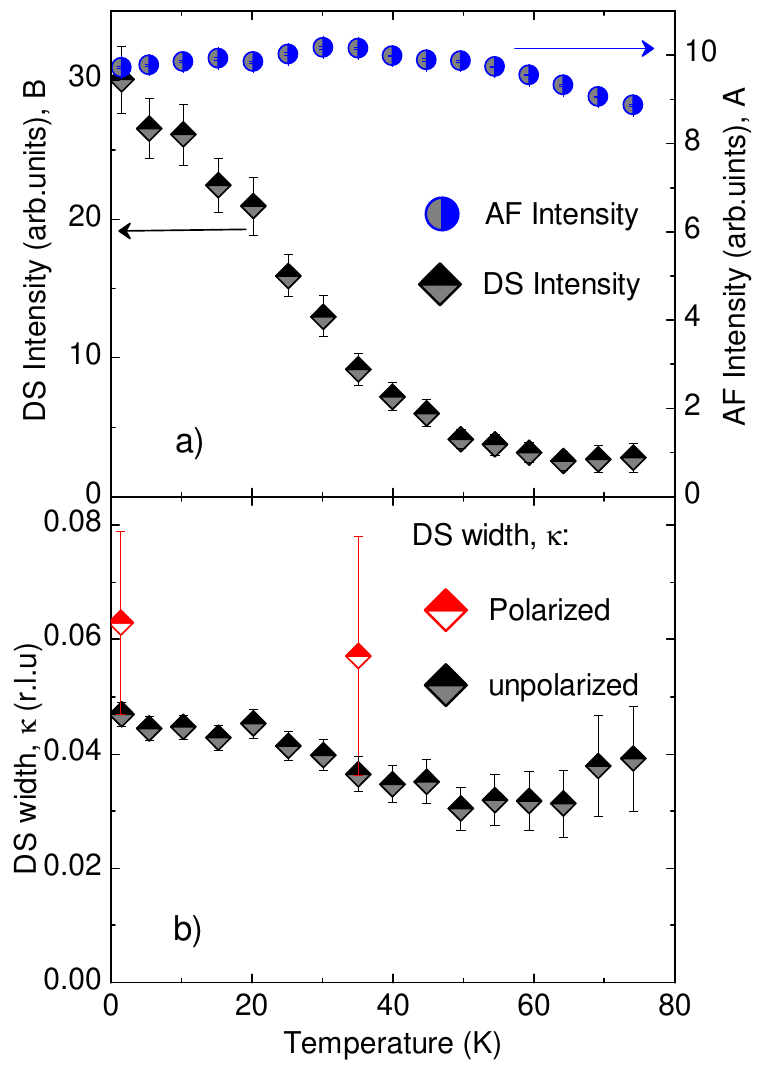}
\caption{a) Integrated intensities of diffuse scattering DS (left Y-axis), AF Bragg peak (right Y-axis) of \pfts, b) The width of neutron DS from \pfts. The values of 
         $\kappa$ inferred from polarized and unpolarized data are in agreement.}
  \label{Fig:pars}
\end{figure}

Having observed these short-range correlations, the first step is to verify their magnetic origin. For this, we make use of polarized neutron scattering with neutron polarization vector parallel to the scattering vector $\mathbf Q$. In this geometry, scattering due to magnetic moments flips the direction of polarization whereas, the polarization state is retained if the scattering is of coherent nuclear origin. In \pfts, the entire AF and DS intensity is observed in the spin-flip channel as shown in Fig.~\ref{Fig:spectra2}. There is nearly no scattering in non spin-flip channel as indicated in the the inset of Fig.~\ref{Fig:spectra2}a. Therefore, we conclude that these short-range correlations are of magnetic origin. To obtain the time scale of the short-range correlations we have measured several inelastic scans through magnetic DS around AF Bragg peak. We find that the fluctuations corresponding to these correlations occur on the timescale longer than 10$^{-11}$~s and therefore refer to quasi-static nature. Having established that the short-range correlations are magnetic and essentially static, we further assumed that they decay exponentially ($e^{-|\vec r |/\xi}$) with distance with some characteristic correlation length $\xi$. An exponential decay implies a lorentzian-squared form for the structure factor of the diffuse scattering. The total structure factor for quasielastic scattering is then written as:

\begin{eqnarray} \label{eq:eq}
S(\mathbf Q,T) & = & S_{AF}(\mathbf Q,T)+S_{DS}(\mathbf Q,T) \label{eq:1}\\
S_{AF}(\mathbf Q,T) &=& A(T) \delta(\mathbf Q-\mathbf Q_{AF}) \label{eq:2}\\
S_{DS}(\mathbf Q,T) &=& B(T)\frac{\kappa}{[{{(\mathbf Q-\mathbf Q_{AF})}^2+\kappa^2]}^2} \label{eq:3}
\end{eqnarray}
\noindent where $S_{AF}(\mathbf Q,T)$) and $S_{DS}(\mathbf Q,T)$ are the Bragg and DS contributions, respectively. $A(T)$, $B(T)$ are temperature dependent integrated intensities of the AF and DS contributions respectively and $\kappa$, inverse correlation length of  DS such that $\kappa = \frac{1}{\sqrt{\sqrt2-1}}\frac{1} {\xi}$.

The best fits of the data to the resolution convoluted scattering function defined above are shown by solid lines in Figs.~\ref{Fig:spectra1}, \ref{Fig:spectra2}.

Performing this analysis at each temperature we get T-dependence of the inverse correlation length $\kappa$, as well as the intensities of the Bragg and diffuse components (Fig.~\ref{Fig:pars}). Intensity of the AF Bragg peak remains nearly unchanged in the covered temperature range as shown in Fig.~\ref{Fig:pars}a. The intensity of DS is weak and remains nearly constant for T$\ge50$~K. However, it increases rapidly upon lowering the temperature below 50~K. The width of the DS remains almost constant as shown in Fig.~\ref{Fig:pars}b, and the corresponding average correlation length is  $\sim$ 10$\pm$2~\AA. 

\subsection{M\"ossbauer spectroscopy}
While neutron scattering provides information on spatial correlations of magnetization, microscopic techniques give a more precise measurement of the distribution of local magnetization. Specifically, in our previous studies of \pfns, a clear proof of coexisting AF and SG order parameters was obtained in  M\"ossbauer spectroscopy experiments. In this work we used a similar  approach for \pfts.

\begin{figure}
\includegraphics[width=1 \columnwidth]{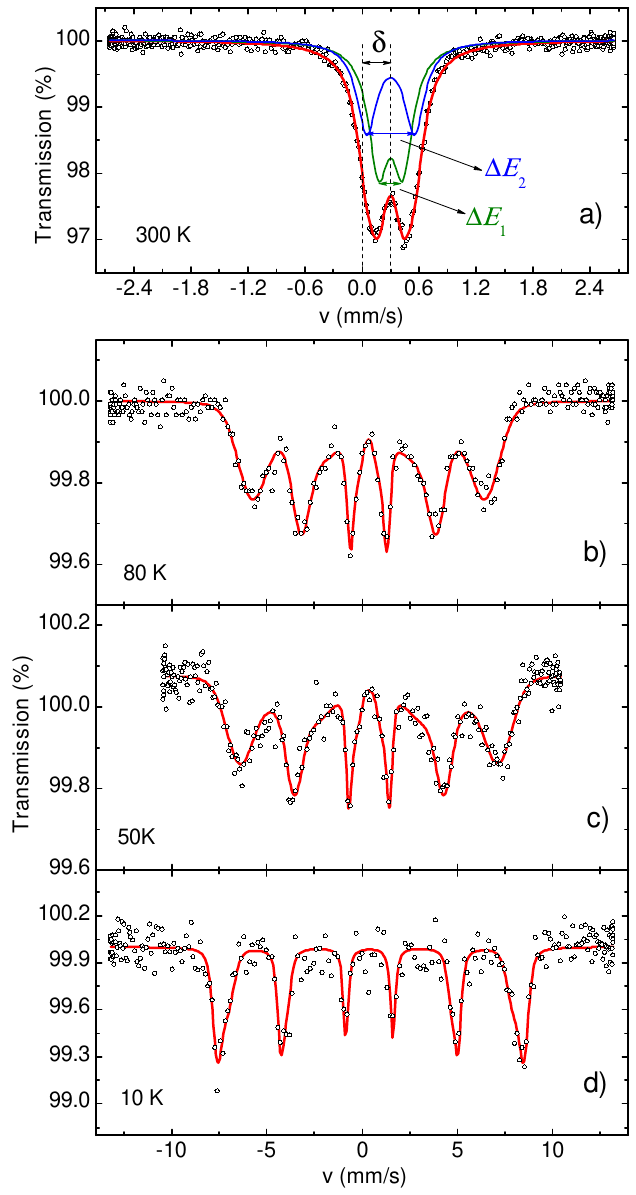}
\caption{M\"{o}ssbauer $^{57}Fe$ absorption spectra of \pfts taken (a) above and (b)-(d) below T$_N$. Red lines are fits to the spectra as described in the text. Two components 
         of the spectrum shown in Fig.~\ref{Fig:Moss1}(a) result from different quadrupole splittings $\Delta E_{1}$, $\Delta E_{2}$ detected in \pfts .}
  \label{Fig:Moss1}
\end{figure}

\begin{figure}
\includegraphics[width=1 \columnwidth]{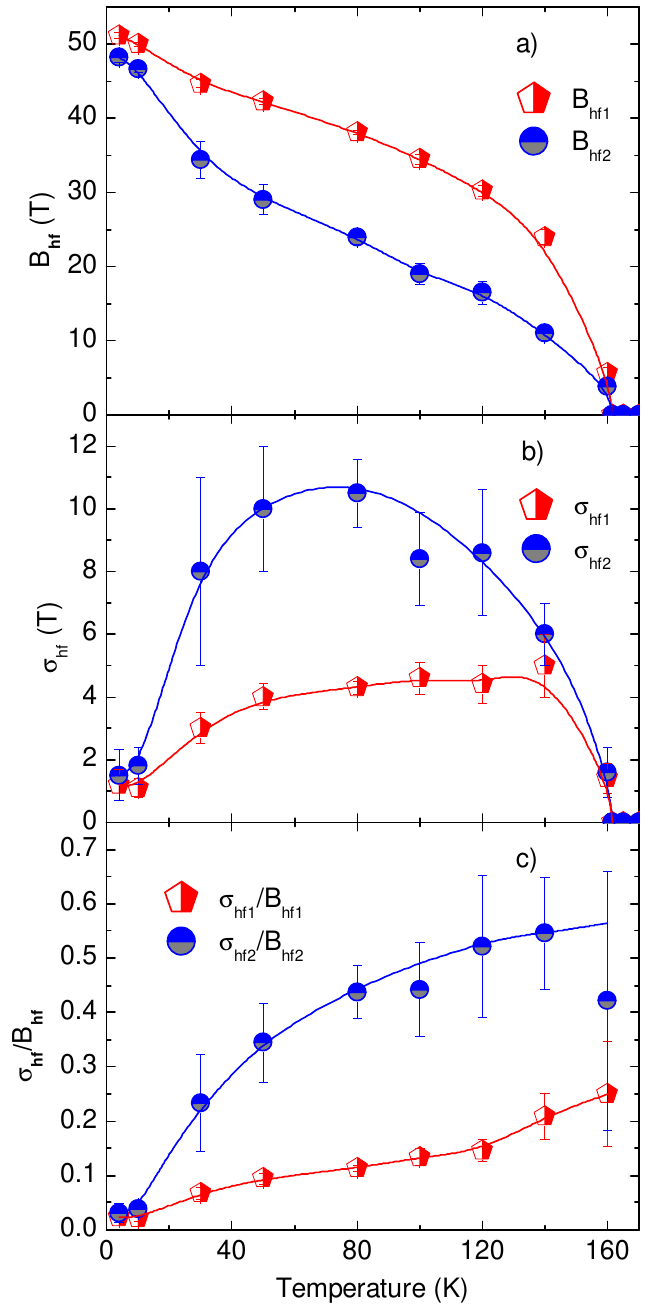}
\caption{Temperature dependence of the model parameters obtained from fits of the M\"{o}ssbauer spectra of \pfts. Here, B$_{hf1}$, B$_{hf2}$  
		are the mean values of hyperfine fields, and $\sigma_{hf1}$, $\sigma_{hf2}$: are the widths of two Gaussian hyperfine field distributions. $\frac{\sigma_{hf}}{B_{hf}}$ gives a measure of relative distribution widths of the hyperfine fields. The lines are guide to the eye.}
  \label{Fig:Moss2}
\end{figure}

\noindent Fig.~\ref{Fig:Moss1} shows representative $^{57}Fe$ M\"{o}ssbauer spectra of \pfts above and at several temperatures below T$_N$. 
The spectrum taken above N\'eel temperature, at T~=~300~K, is shown in Fig.~\ref{Fig:Moss1}a. It exhibits a doublet structure with its center of weight shifted to $\delta$~=~0.301(2)~mm/s (with respect to energy of $^{57}CoRh$ source kept at room temperature). This isomer shift is mainly determined by the chemical shift and clearly corresponds to trivalent state of Fe. The doublet shape in the spectrum arises due to quadrupolar splitting produced by a non-spherical charge distribution around Fe$^{3+}$ ion. Attempts to fit the spectrum of Fig.~\ref{Fig:Moss1}a assuming one electric field gradient acting at Fe$^{3+}$,  lead to unsatisfactory description with high $\chi^2$ and an excessive line broadening. Instead, an adequate fit  is achieved by assuming two contributions to the quadrupolar splittings $\Delta E_{1}=0.26(1)$, $\Delta E_{2}=0.50(2)$ as shown in Fig.~\ref{Fig:Moss1}a. Having two quadrupolar splittings at a single iron site is not surprising, as \pfts is already in the ferroelectric phase at T~=~300~K and an additional deviation from the spherical approximation can be induced by different formal charges of disordered Fe$^{3+}$ and Ta$^{5+}$ ions.  The ratio of areas of two components in the spectra is found to be 0.57(3):0.43(3).

Below the N\'eel temperature (Figs.~\ref{Fig:Moss1}b-d), the M\"{o}ssbauer spectra of \pfts exhibit a sextet pattern. This is to be interpreted as a direct evidence of Zeeman splitting of $^{57}Fe$ nuclear levels induced by the spontaneous magnetization of the material. No central (unsplit doublet) contribution similar to the spectrum shown in Fig.~\ref{Fig:Moss1}a could be detected. This rules out the existence of remaining paramagnetic clusters~\cite{Nomura1968} below T$_{N}$ and clearly indicates that all Fe$^{3+}$ in \pfts are involved in creating AF long range order. The nuclear-quadrupole splitting which was the dominant effect of the spectra in the paramagnetic state is not detected below T$_{N}$. This points to a wide distribution of the angles between the iron spins and the local axes of the electric field gradient tensor in the AF state. The sextet itself is noticeably broadened at higher temperatures. At base temperature this broadening nearly vanishes and sharp absorption lines are observed, indicating static and rather uniform local fields around Fe$^{3+}$ sites. 

To obtain more quantitative information from M\"{o}ssbauer data, the spectra taken below T$_{N}$ were fitted with hyperfine fields having Gaussian distributions (see Fig.~\ref{Fig:Moss1}(b)-(d)). A consistent fit for all the spectra can be obtained with a single isomer shift for Fe$^{3+}$ assuming two distributions of the hyperfine fields in the same ratio as was inferred from the data in paramagnetic phase. Fig.~\ref{Fig:Moss2}a shows the temperature dependences of the amplitudes of both hyperfine fields B$_{hf1}$, B$_{hf2}$. B$_{hf}$ for the two contributions increases smoothly on cooling from T$_{N}$ down to $\sim$50~K. Below 50~K a faster increase of B$_{hf}$ is observed, and both finally reach essentially the same value of $\sim$50~T at 4~K. This value is very close to the saturated hyperfine field in other Fe containing oxides~\cite{Chillal2013},\cite{Shirane1962} and thus suggests that Fe$^{3+}$ in \pfts recovers a full moment at base temperature. The strong temperature dependence of widths $\sigma_{hf1}$, $\sigma_{hf2}$  (see Fig.~\ref{Fig:Moss2}b) suggests that the origin of observed distributions of the hyperfine fields is due to dynamic fluctuations. This dynamic nature is seen even better through the temperature evolution of the ratio $\frac{\sigma_{hf}}{B_{hf}}$ shown in Fig.~\ref{Fig:Moss2}c.  These relative widths of B$_{hf1}$ and B$_{hf2}$ decrease monotonically as temperature is lowered. For both contributions it reaches the same and vanishingly small value at base temperature.  This implies negligible contribution from static fluctuations to the broadening of the M\"{o}ssbauer spectra.

The features of the M\"{o}ssbauer spectra below T$_{SG}$ unambiguously confirm a homogeneous environment for all Fe$^{3+}$ ions in the system, and a uniformity of the {\it magnitude} of the moments at saturation. Since neutron scattering demonstrated an {\it increase} of short-range correlations in this regime, we conclude that the disorder at low temperature is purely orientational. In addition, the slowing down of hyperfine field fluctuations results in a rapid increase of local Fe moment below 50~K. In contrast, the ordered magnetic moment seen by neutrons remains nearly unchanged (possibly reduces slightly) below this temperature. A clear difference in the behavior of local and staggered magnetic moments below $\sim50$~K indicates canting of Fe moments. A similar behavior was observed in amorphous metallic glasses Fe-Mn, Fe-Zr, Au-Fe and was ascribed to the spin canting~\cite{Ryan2003,Ren1995,Campbell1992}. This further validates our interpretation of the observations for \pfts. Therefore, these results for \pfts support the same scenario for the AFSG phase, that we previously suggested for PFN.

\section{Summary}

In summary, we have shown that the magnetic properties of \pfts are essentially identical to those of \pfns. \pfts undergoes only two magnetic phase transitions: one of which is AF (T$_{N}\sim$153~K) and the other is SG (T$_{SG}\sim$10~K) to the AFSG ground state, contrary to the previous claims of two N\'eel temperatures, namely T$_{N1}\sim$160~K and T$_{N2}\sim$48~K~\cite{Lampis2004,Martinez2010}. Our experiments allow us to associate the latter temperature with the enhancement of magnetic short-range correlations in the sample which are further developed as $T_{SG}$ is approached. These correlations are revealed by an increase of the integrated intensity of DS scattering (Fig.~\ref{Fig:pars}a), a similar feature also observed in \pfns. In \pfts, the measured correlation length 10~$\pm2~\AA$ corresponds to nearly two lattice constants which clearly suggests that magnetic interactions beyond first nearest-neighbor are important.\footnote{It should be noted that the correlation length calculated for \pfns by {\it Rotaru} et. al~\cite{Gelu2009} would be $\sim$10~$\AA$ if a 4D convolution is performed and a squared Lorentzian is used to fit DS.} 

Through temperature evolution of M\"ossbauer spectra, we observe that the fluctuations in hyperfine field slow down as T$_{SG}$ is approached, in accordance with increasing intensity of short-range correlations. 

To conclude, the nature of the magnetic phases and specifically the microscopic coexistence of long range AF and orientational SG order are a common feature of stoichiometric disordered PbFe$_{1/2}$B$_{1/2}$O$_{3}$ perovskites. A non-isovalent dilution on the B-site affects the LRO as well as the ground state~\cite{Laguta2013}. We establish that as long as the dilution is isovalent, the non-magnetic ions do not produce a strong effect on the magnetic phase transitions in this family.

\bibliographystyle{apsrev4-1}
\bibliography{PFT}{}
\end{document}